\documentclass[10pt,floatfix,%
twocolumn,
amsmath,amssymb,
aps,
superscriptaddress,
prapplied,
]{revtex4-2}

\usepackage[svgnames,dvipsnames]{xcolor}
\usepackage{url}
\usepackage{graphicx}
\graphicspath{{.}{figs/}{../figs/}}
\usepackage{afterpage}
\usepackage{dcolumn}
\usepackage{bm}
\usepackage{tikz}
\usetikzlibrary{quantikz2,positioning,intersections,angles,quotes}
\usepackage{adjustbox}
\usepackage[caption=false]{subfig}
\usepackage{comment}
\usepackage{physics}
\usepackage{mathtools}

\usepackage[colorlinks,citecolor=DarkGreen,linkcolor=FireBrick,linktocpage,unicode,hypertexnames=false]{hyperref}
\usepackage[capitalize]{cleveref}
\crefname{enumi}{}{}
\crefname{section}{Sec.}{Sec.}
\usepackage{autonum}
\usepackage{booktabs}
\usepackage{multirow}
\usepackage{algorithm}
\usepackage{algpseudocode}

\begin{document}
\title{Data-driven adaptive quantum error mitigation for probability distribution}

\author{Rion Shimazu}
\affiliation{NTT Computer and Data Science Laboratories, NTT Inc., Musashino 180-8585, Japan}

\author{Suguru Endo}
\affiliation{NTT Computer and Data Science Laboratories, NTT Inc., Musashino 180-8585, Japan}
\affiliation{NTT Research Center for Theoretical Quantum Information, NTT Inc., 3-1 Morinosato Wakanomiya, Atsugi, Kanagawa, 243-0198, Japan}

\author{Shigeo Hakkaku}
\affiliation{NTT Computer and Data Science Laboratories, NTT Inc., Musashino 180-8585, Japan}
\affiliation{NTT Research Center for Theoretical Quantum Information, NTT Inc., 3-1 Morinosato Wakanomiya, Atsugi, Kanagawa, 243-0198, Japan}

\author{Shinobu Saito}
\email{shinobu.saito@ntt.com}
\affiliation{NTT Computer and Data Science Laboratories, NTT Inc., Musashino 180-8585, Japan}

\begin{abstract}
Quantum error mitigation (QEM) has been proposed as a class of hardware-friendly error suppression techniques. While QEM has been primarily studied for mitigating errors in the estimation of expectation values of observables, recent works have explored its application to estimating noiseless probability distributions.
In this work, we propose two protocols to improve the accuracy of QEM for probability distributions, inspired by techniques in software engineering. The first is the N-version programming method, which compares probability distributions obtained via different QEM strategies and excludes the outlier distribution, certifying the feasibility of the error-mitigated distributions. The second is a consistency-based method for selecting an appropriate extrapolation strategy. Specifically, we prepare $K$ data points at different error rates, choose $L<K$ of them for extrapolation, and evaluate error-mitigated results for all $\binom{K}{L}$ possible choices. We then select the extrapolation method that yields the smallest variance in the error-mitigated results. This procedure can also be applied bitstring-wise, enabling adaptive error mitigation for each probability in the distribution.

\end{abstract}
\maketitle
\section{Introduction}
Near-term quantum computing with noisy intermediate-scale quantum (NISQ) or early fault-tolerant quantum computers is error-prone due to the limited functionality of quantum error correction (QEC)~\cite{devitt2013quantum,lidar2013quantum,preskill2018quantum,preskill2025beyond,eisert2025mind}. Due to its hardware efficiency, quantum error mitigation (QEM) is an alternative to QEC~\cite{cai2023quantum,endoHybridQuantumClassicalAlgorithms2021,endoPracticalQuantumError2018}. QEM post-processes the measurement outcomes from quantum computers to estimate the error-mitigated result at the cost of more repetitions on quantum computers. 

QEM has been considered a powerful tool for high-accuracy quantum computing, and various QEM methods have been proposed so far. For example, zero-noise extrapolation (ZNE) estimates the error-free result by extrapolating from erroneous and error-boosted results~\cite{temmeErrorMitigationShortDepth2017,li2017efficient,endoPracticalQuantumError2018}. Although the ZNE method is hardware-efficient and has shown significant experimental success in improving computational accuracy \cite{kimEvidenceUtilityQuantum2023}, it remains a heuristic approach without a formal guarantee of accuracy. Probabilistic error cancellation (PEC) tries to cancel the effect of noise by utilizing characterized noise information obtained in advance \cite{temmeErrorMitigationShortDepth2017,endoPracticalQuantumError2018}. However, noise characterization can never be perfect, which causes an inevitable bias in error-mitigated results. Since the accuracy and bias of these QEM methods are unknown in general, there is no established way for QEM users to select the most suitable QEM method for their use cases. This poses the risk that they may inadvertently employ an unsuitable QEM method with severely limited accuracy.

Traditionally, QEM techniques have primarily been employed to estimate the expectation values of observables computed from the output of quantum circuits~\cite{endoHybridQuantumClassicalAlgorithms2021,cai2023quantum}. However, recent research has begun to explore QEM methods aimed at reconstructing the \textit{entire output distribution} of the output of noisy quantum circuits \cite{liu2025quantum, muqeetQUIETToolforSampling-BasedQuantumNoiseErrorMitigation}. For instance, the method proposed in Ref.~\cite{liu2025quantum} takes this approach by estimating the error-mitigated output distribution based on the noisy measurement data, and then sampling from the reconstructed distribution. The authors of Ref.~\cite{liu2025quantum} benchmarked their method on a toy model of the quantum phase estimation task \cite{nielsenQuantumComputationQuantum2012}, demonstrating its applicability and discussing its potential advantages. Notably, such QEM techniques targeting output distributions are not limited to quantum phase estimation, but are also applicable to other distribution-based quantum tasks such as quantum-selected configuration interaction \cite{kanno2023quantumselectedconfigurationinteractionclassical}. In fact, QEM was employed in the experiments of the quantum-selected configuration interaction, as reported in Ref.~\cite{Nakagawa2024adaptQSCI}. Therefore, the methods proposed in Refs.~\cite{kimEvidenceUtilityQuantum2023,temmeErrorMitigationShortDepth2017} represent a significant broadening of the scope of QEMs. However, as with conventional QEM approaches, selecting an appropriate error mitigation strategy remains a key challenge in practice.

\begin{figure}[tb]
    \centering
    \includegraphics[width=1.0\linewidth]{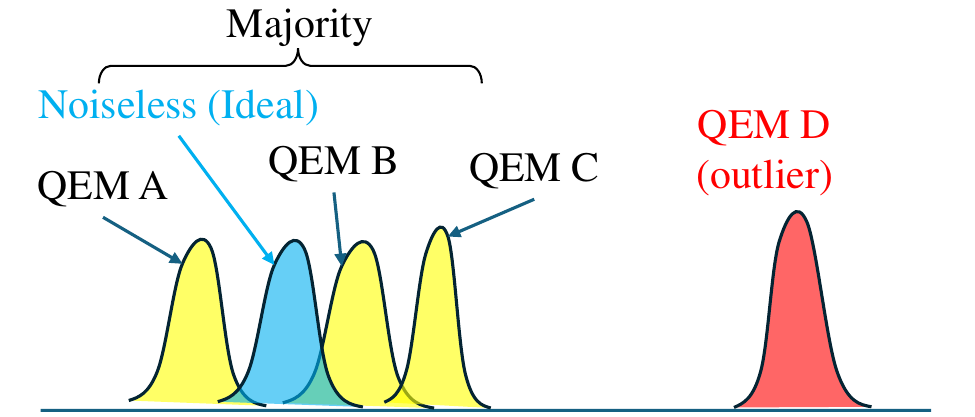}
    \caption{The N-version programming approach calculates the distances among the distributions of the given QEM methods (QEMs A to D in this figure) and effectively identifies the method that deviates most from the majority (in this case, QEM D).}
    \label{fig: N-version programming method}
\end{figure}

Here, we propose two methods to systematically and adaptively select the feasible QEM strategy, mainly for the error-mitigated probability distribution. The first method is inspired by the $N$-version programming used in the classical software engineering field~\cite{NVP1995}.
The conceptual figure is shown in Fig.~\ref{fig: N-version programming method}.
For instance, in Ref.~\cite{SaitoRAQuS2024}, the $N$-version of programming certifies the computation outcomes by comparing the probability distributions obtained from similar computations, e.g., probability distributions corresponding to solving the same problem using different algorithms. Because we can prepare several error-mitigated distributions through Ref.~\cite{liu2025quantum} via different applications of QEM strategies, we can compare these error-mitigated distributions with each other and choose the accurate probability distribution based on certain metrics. Here, we select the probability distribution that has the smallest total variational distance (TVD) to other distributions, because a large TVD indicates that the probability distribution is highly likely to be atypical. 

\begin{figure}[tb]
    \centering
    \includegraphics[width=\linewidth]{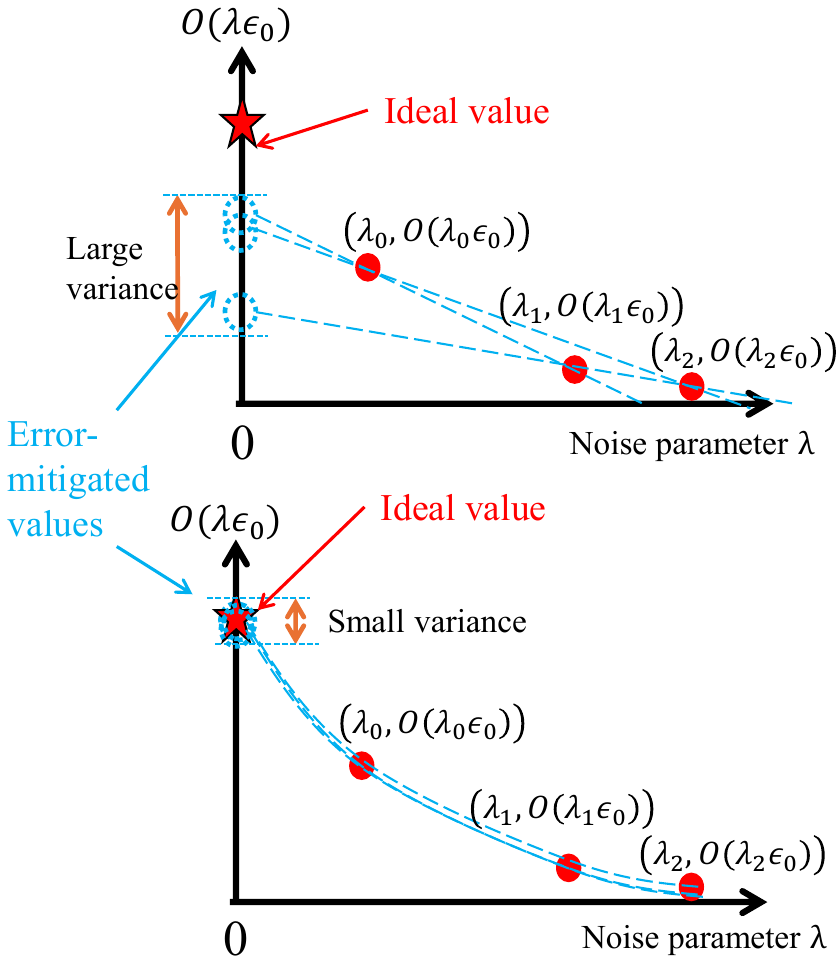}
    \caption{The consistency-based method. In this approach, we hypothesize that when the optimal QEM method is employed, the mitigation results remain almost unchanged regardless of which data points are used. Under this assumption, the method that yields the smallest variance in QEM results across different data points is regarded as the most ``consistent'', and thus identified as the optimal one.}
    \label{fig: consistency-based method}
\end{figure}

The second method selects a suitable QEM method by comparing the output obtained using different QEM strategies to check the consistency of each QEM strategy. 
This method is also inspired by MAPE-K, commonly used to construct self-adaptive software systems~\cite{computing2006architectural}. MAPE-K defines four functions (\textit{M}onitor, \textit{A}nalyze, \textit{P}lan, \textit{E}xecute) and a \textit{K}nowledge base that manages data shared between them. It enables systems to adapt to changing conditions and maintain optimal performance. By employing the concept of MAPE-K (i.e., self-adaptive), we can plan and execute the selection of the most suitable QEM method through monitoring and analysis of multiple different error rates (i.e., changing conditions).
The conceptual figure for our protocol is shown in Fig.~\ref{fig: consistency-based method}. First, we prepare data points for several error rates, choosing only a subset for extrapolation. We then compare the results of different extrapolation strategies for all possible combinations of the chosen data points. Next, we select the extrapolation strategy that yields the lowest variance in the error-mitigated outcomes. While we can perform this method to improve the estimation of the single-observable expectation value, we can apply this method to each measurement probability for the computation basis measurement, improving the accuracy of the overall error-mitigated probability distribution.

We verify the performance of our methods through numerical simulations using a Trotter-based Hamiltonian simulation algorithm for a transverse Ising Hamiltonian. For the N-version programming method, we compare the error-mitigated distributions obtained from extrapolation strategy including that provided by the Mitiq~\cite{mitiq2022Quantum} QEM software package, and confirmed that we can systematically exclude the error-mitigated distribution that deviates significantly from others and the ideal distribution. In addition, when applying the consistency method to each bin of the probability distribution, it selects the most suitable QEM strategy in the dominant cases, yielding a naturally error-mitigated distribution with the smallest distance from the ideal distribution. 

\section{Quantum error mitigation for probability distribution}

Here, we describe the recently introduced quantum error mitigation for probability distribution presented in Ref.~\cite{liu2025quantum}. Note that the expectation value of each POVM element $\Pi_z = \ket{z}\bra{z}$ is equivalent to the probability $p_z$ of obtaining the bitstring $z$, i.e., $p_z= \text{Tr}[\rho \Pi_z]$ for a density matrix $\rho$ representing the quantum state. We can interpret this method as mitigating the expectation value of each POVM element. Ref.~\cite{liu2025quantum} mainly considers QEM methods using the linear ansatz~\cite{cai2021practical}:
\begin{equation}
\rho_\text{QEM}= \sum_k c_k \rho_k,
\end{equation}
where $c_k \in \mathbb{R}$ and $\rho_k$ is a noisy state. The error-mitigated state $\rho_\text{QEM}$ obtained via extrapolation and probabilistic error cancellation can be explained with the linear ansatz QEM state. Then, each error-mitigated probability $p_z^\text{QEM}$ can be represented as 
\begin{equation}
\begin{aligned}
p_z^\text{QEM}
&= \text{Tr}\bqty{\Pi_z\rho_\text{QEM}}\\
&= \sum_k c_k \text{Tr}[\Pi_z \rho_k] \\
&= \Gamma \sum_k p_k \text{sgn}(c_k) \text{Tr}[\Pi_z \rho_k],
\label{Eq: Linear}
\end{aligned}
\end{equation}
where $\Gamma=\sum_k |c_k|$ and $p_k=\abs{c_k}/\Gamma$. The third line in Eq.~\eqref{Eq: Linear} allows for the Monte-Carlo implementation of the error-mitigated probability $p_z^\text{QEM}$. We randomly generate the state $\rho_k$ with probability $p_k$ and measure the state with the computational basis. We define the random variable 
\begin{equation}
\hat{\mu}_z = \text{sgn}(c_k) \hat{m}_z,
\end{equation}
where $\hat{m}_z=1$ when the measured result is $z$ and otherwise $0$. Then, we can show
\begin{equation}
p_z^\text{QEM}=\Gamma \mathbb{E} [\hat{\mu}_z]. 
\end{equation}
Because of $[\Pi_z, \Pi_z']=0 $ $\forall z, z'$, the computational basis measurement allows for the simultaneous estimation of the overall probability distribution $\{p_z\}_z$, i.e., when $\hat{m}_z=1$, $\hat{m}_{z'}=0$ for $\forall z' \neq z$. Then, Ref.~\cite{liu2025quantum} shows 
\begin{equation}
\sum_z \mathrm{Var}[p_z^\text{QEM}] \leq \frac{\Gamma^2}{N_\text{meas}}
\end{equation}
for the number of measurements $N_\text{meas}$. This indicates that the sampling overhead of the probability distribution estimation is comparable to that of a single-observable estimation.

\section{Quantum error mitigation for probability distribution without Monte-Carlo and beyond linear ansatz}

However, the Monte-Carlo implementation of QEM methods is not hardware-friendly because it requires changing the structure of the quantum circuit for each execution. Here, we consider simply estimating the error-mitigated probability $p_z^\text{QEM}$ as in the first line of Eq. \eqref{Eq: Linear}, i.e., $p_z^\text{QEM}=\sum_k c_k \text{Tr}[\Pi_z \rho_k] = \sum_k c_k p_z^{(k)} $. In this case, the variance of the error-mitigated probability can be described as
\begin{equation}
\begin{aligned}
\mathrm{Var}[p_z^\text{QEM}]&= \sum_k c_k^2 \text{Var}[p_z^{(k)} ] \\
&= \sum_k c_k^2 \frac{p_z^{(k)} - p_z^{(k) 2} }{N_\text{meas} ^{(k) }}, 
\end{aligned}
\end{equation}
which results in
\begin{equation}
\begin{aligned}
\sum_z \text{Var}[p_z^\text{QEM}]&= \sum_k c_k^2 \frac{1- \sum_z p_z^{(k) 2}}{N_\text{meas} ^{(k) }}  \\
&\leq \sum_k \frac{ c_k^2}{N_\text{meas} ^{(k) }}, 
\end{aligned}
\end{equation}
where $N_\text{meas} ^{(k) }$ is the number of samples assigned to measure the noisy state $\rho_k$, satisfying $\sum_k N_\text{meas} ^{(k) }= N_\text{meas} $. 

Compared with the Monte-Carlo implementation, the sampling overhead generally becomes worse as $\sum_k c_k^2/N_\text{meas} ^{(k) } \geq  (\sum_k |c_k|)^2/N_\text{meas}  $. Note that, by optimizing the sample number allocation, the two sampling overheads become equivalent. Therefore, when the number of erroneous states in the linear ansatz in Eq.~\eqref{Eq: Linear} is not significant; the straightforward QEM strategy works as an alternative method to the Monte-Carlo method. 

Furthermore, when we do not resort to the Monte-Carlo implementation, we are not restricted to the linear ansatz in Eq.~\eqref{Eq: Linear}. For example, while the exponential extrapolation method cannot be described by the linear ansatz, it shows excellent performance in actual large-scale experiments. Although the exponential extrapolation method is not applicable when $p_z=0$ due to division, it is desirable to incorporate this method when possible.

\section{Proposed methods}
Many QEM methods have been proposed, but the accuracy of those methods varies significantly depending on factors such as the noise characteristics of quantum computers or the structure of quantum circuits. Therefore, it is important to select the method that maximizes accuracy according to the specific use case. However, selecting the best method is difficult because the effect of noise characteristics and circuit structure on the accuracy of QEMs is unknown in general. This may make QEM users choose an unsuitable QEM method for their use cases, and users would be left with inaccurate results of QEMs. To overcome the challenge, we propose two methods that utilize multiple QEM methods to produce high-accuracy results from quantum computers: the N-version programming method and the consistency-based method.

\subsection{N-version programming method}\label{sec: N-version programming method}

The first method is the N-version programming method. N-version programming is a concept that is frequently employed in the field of software engineering to obtain an error-free result~\cite{NVP1995}. The basic concept involves solving the same problem with different setups, such as various algorithms and environments, and then comparing the outcomes from each setup to extract the feasible ones. The celebrated example of the N-version of programming is to compute the distance of each probability distribution and reject the distribution with the largest distance from other distributions to avoid atypical results~\cite{SaitoRAQuS2024}.

Here, we propose applying the N-version programming methods to the error-mitigated probability distributions. 
The schematic figure of this method is shown in Fig.~\ref{fig: N-version programming method}. 
In our case, the ``N-versions'' are obtained through the applications of different QEM strategies. For example, we prepare three types of error-mitigated probability distributions via linear extrapolation, second-order Richardson extrapolation, and exponential extrapolation from noisy probability distributions. For extrapolation QEM methods, we can prepare the ``N-versions" only via classical processing of noisy probability distributions prepared in advance. After we prepare $N$ instances of error-mitigated probability distributions, we compute the TVD for all the $\binom{N}{2}$ pairs of error-mitigated probability distributions. Then, denoting each error-mitigated distribution $\{q_k\}_{k=1}^{N_\text{QEM}}$ with $N_\text{QEM}$ being the number of error-mitigated distributions, we compute TVDs $D(q_k, q_k')$ for $\forall k,k'$, and select $k$ such that $\sum_{k'=1}^{N_\text{QEM}} D(q_k, q_k')$ is minimized.

\subsection{Consistency-based  method}\label{sec: consistency-based method}

In the second method, we select the most consistent QEM method by comparing outcomes from different extrapolation QEM methods. The conceptual figure is shown in Fig.~\ref{fig: consistency-based method}. We first prepare the noisy expectation values for $K$ different noise parameters, denoted as $\{ \langle O(\lambda_k \varepsilon_0) \rangle \}_{k=0}^{K-1}$ for noise stretch factor $1=\lambda_0 < \lambda_1 < \lambda_2 ...< \lambda_{K-1}$ and the original error rate $\varepsilon_0$. Then, we select $L<K$ data points to apply the $L$-point extrapolation method. Then, by performing the extrapolation for $\binom{K}{L}$ combinations of erroneous data points, we can prepare $\beta=\binom{K}{L}$ error-mitigated values, denoted as $\{a_j\}_{j=1}^\beta$. Then we compute the variance of the error-mitigated values $\{a_j\}_{j=1}^\beta$. Note that a small variance of the error-mitigated values indicates that the extrapolation strategy yields a consistent error-mitigated result. We perform the same procedures for other extrapolation strategies, and choose the QEM strategy with the smallest variance as the most consistent extrapolation curve. While the extrapolation method is inherently heuristic and generally lacks a theoretical guarantee of accuracy, our approach enables us to establish a certain degree of performance guarantee. We can also apply this method in a bit-string-wise manner to each probability distribution $\{p_z^\text{QEM}\}_z$, e.g., we apply a linear extrapolation for $z=01$ and an exponential extrapolation for $z=10$, which is determined by the consistency evaluation. 

\section{Numerical Experiments}\label{sec: Numerical Experiments}

\subsection{Experimental setup and procedure}\label{sec: Experimental setup and procedure}
\begin{figure*}[tb]
    \centering
    \includegraphics[width=0.7\linewidth]{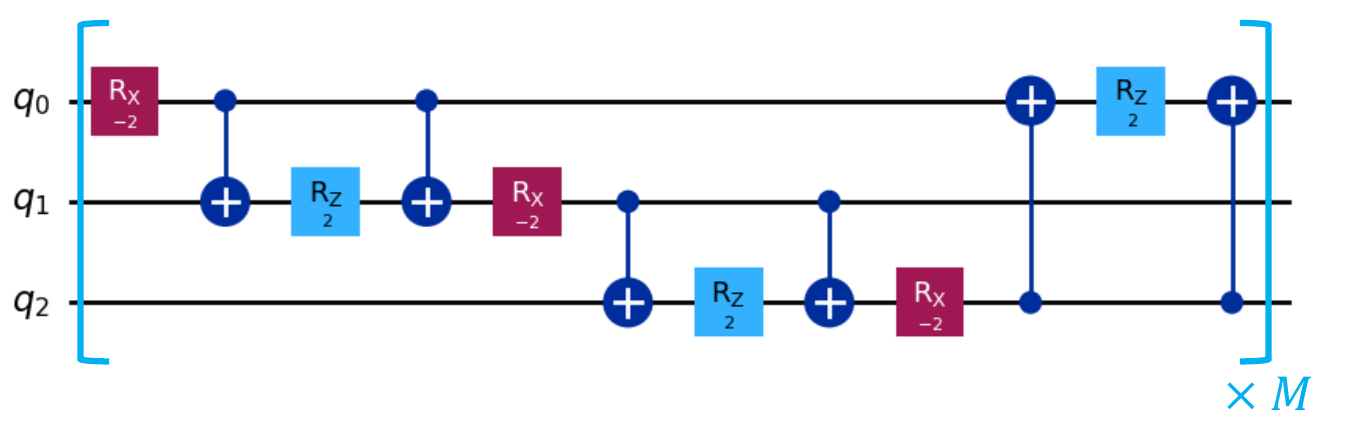}
    \caption{An example of a 3-qubit Trotterized circuit for the one-dimensional transverse-field Ising model. A single block composed of RX, RZ, and CNOT gates is repeatedly applied for $M$ times. We used Qiskit \cite{qiskit2024} to draw the circuit.}
    \label{fig: example_of_Trotterized_circuit}
\end{figure*}
\begin{table}[tb]
    \centering
    \begin{tabular}{|l|l|l|}
    \hline
        \textbf{Parameter} & \textbf{Description} & \textbf{Value(s)} \\ \hline \hline 
        $N_Q$ & Numer of qubits  & =10 \\ \hline 
        \multirow{2}{*}{$N_\textrm{meas}$} & Number of measurements  & \multirow{2}{*}{=5000} \\ 
        & for each circuit & \\ \hline 
        \multirow{2}{*}{$t$} & Duration of  & \multirow{2}{*}{=1} \\ 
        & the time evolution & \\ \hline 
        $J$ & Coefficients of  &   \multirow{2}{*}{$\in\{1,2,\cdots,10\}$} \\ \cline{1-1}
        $B$ & the Hamiltonian &    \\ \hline
        $\lambda$ &  Scale factor &  $\in\{3,5\}$  \\ \hline
    \end{tabular} \\
    \caption{The common parameters used in the experiments. }
    \label{tab: parameters used in the experiments}
\end{table}
For each of the two proposed methods, we conduct numerical experiments using a quantum circuit simulator. In our numerical experiments, we consider the time evolution of a physical model described by the following one-dimensional transverse-field Ising Hamiltonian:
\begin{align}
    H_\text{TFI} = J \sum_{j=1}^{N_Q} Z_j Z_{j+1} + B\sum_{j=1}^{N_Q} X_j,
\end{align}
where $Z_i$($X_i$) is the Pauli Z(X) operator on the $i$-th site. For this Hamiltonian, the time-evolution operator over a period $t$ is expressed as $e^{-iH_\text{TFI}t}$. Here, in order to represent this time-evolution operator as a quantum circuit, we perform the following first-order Trotterization:
\begin{align}
    e^{-iH_\text{TFI}t}&  \approx U_T,
\end{align}
where
\begin{align}
    U_T \equiv  \Bigg\{ & \exp  \Bigg(i\sum_{j=1}^{N_Q} Z_j Z_{j+1}\frac{t}{M}\Bigg) \notag \\
        & \times \exp\Bigg(-i\sum_{j=1}^{N_Q} X_i \frac{t}{M}\Bigg) \Bigg\}^M,
\end{align}
with $M$ the Trotter number. See Sec.~\ref{sec: Trotterization} for the details of the first-order Trotterization. For the simulation of the circuit $U_T$, we employ Qiskit \cite{qiskit2024}. Figure~\ref{fig: example_of_Trotterized_circuit} shows an example of a 3-qubit Trotterized circuit implementing the unitary transformation $U_T$. The circuit $U_T$ can be expressed using the rotation gate and CNOT gate classes provided by Qiskit. The simulations are performed using the backend named \texttt{FakeMarrakesh}, which simulates the noise characteristics of IBM's quantum hardware. Other software used in the simulations is summarized in Tab.~\ref{tab: software used for the experiments.} in Sec.~\ref{sec: Software table}.

The parameters used in our experiments are shown in Tab.~\ref{tab: parameters used in the experiments}. In this numerical experiment, the coefficients $J$ and $B$ of the Hamiltonian $H_\text{TFI}$ are varied from 1 to 10, resulting in a total of $10\times10=100$ combinations of $(J,B)$. For each pair of $J$ and $B$, we first simulate the unitary operation $U_T$ and compute both the output distribution $\{p_z\}_z$. Subsequently, we construct a modified circuit $U_{T,\lambda}$ whose error is amplified by a scale factor $\lambda$, and simulate the corresponding output distribution $\{p_{z,\lambda}\}_z$. Note that these distributions are later used for performing quantum error mitigation. The procedure for amplifying the error is presented in Sec.~\ref{sec: Error amplification}. 

After obtaining the distributions $\{p_z\}_z$ and $\{p_{z,\lambda}\}_{z,\lambda}$, we apply the QEM methods and the proposed methods using these results and obtain the error-mitigated distributions $\{p_z^\text{QEM}\}_z$. In this experiment, we adopt extrapolation-based QEM methods \cite{endoHybridQuantumClassicalAlgorithms2021, endoMitigatingAlgorithmicErrors2019}. Specifically, we employ linear extrapolation, second-order Richardson extrapolation, exponential extrapolation, and polynomial-exponential extrapolation methods. The first three methods estimate the error-mitigated probability distribution $\{p_z^\text{QEM}\}_z$ according to the following equations:

\begin{description}
    \item[Linear] 
    \begin{equation}
        p_z^\text{QEM} = \frac{3p_z - p_{z,3}}{2}. \notag
    \end{equation}
    \item[Second-order Richardson]
    \begin{equation}
        p_z^\text{QEM} =  \sum_{\lambda\in \{1,3,5\}} C_{\lambda} p_{z,\lambda}, \notag 
    \end{equation}
    where $p_{z,1}=p_z$ and 
    \begin{equation}
        C_{\lambda} = \prod_{\lambda,\lambda'\in\{1,3,5\},\;  \lambda \neq \lambda'} \frac{\lambda'}{\lambda' - \lambda}. \notag 
    \end{equation}
    \item[Exponential]
    \begin{equation}
        p_z^\text{QEM} =  (p_z)^{3/2}  (p_{z,3})^{-1/2}. \notag 
    \end{equation}    
\end{description}

In the fourth method, $p_z^\text{QEM}$ is estimated numerically via nonlinear regression for the polynomial-exponential extrapolation function, rather than by an analytical formula. The employed nonlinear regression model is as follows:

\begin{description}
\item[Polynomial-Exponential Extrapolation]
    \begin{equation}
        p_z^\text{QEM} =\theta_0 \exp(\theta_1\lambda + \theta_2 \lambda^2). \notag 
    \end{equation}

\end{description}
 Here, $\theta_0,\theta_1,\theta_2 \in \mathbb{R}$ are the parameters determined by the numerical fitting.
This model is among those supported by the Mitiq package \cite{mitiq2022Quantum}. We employ Mitiq's \texttt{PolyExpFactory} class to perform the numerical fitting.

Note that, regardless of the QEM method, the QEM procedure is applied independently for each $z$ only if $p_z$ or any of the values in the set $\{p_{z,\lambda}\}_{z,\lambda}$ is nonzero. If $p_z$ and all values in $\{p_{z,\lambda}\}_{z,\lambda}$ are zero for a given $z$, the corresponding error-mitigated probability $p_z^\text{QEM}$ is set to zero. Then, the TVD between the error-mitigated distributions generated by each QEM method and the ideal distribution is calculated, and the methods are ranked by their TVD values. The procedure described above is performed for all combinations of $J$ and $B$, with each taking values from 1 to 10, yielding a total of 100 runs.

\subsection{Numerical results}\label{sec: Experimental results}

\begin{figure}[b]
    \centering
    \includegraphics[width=1.0\linewidth]{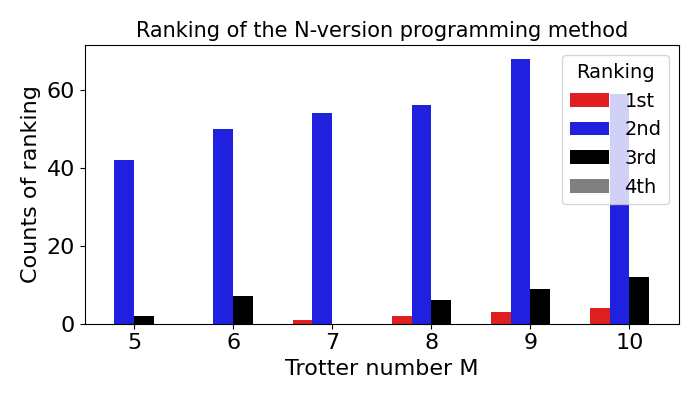}
    \caption{Count of the N-version-programming method’s ranks among QEM methods sorted by the corresponding TVDs. The N-version programming method is never ranked 4th for any parameter $M$, indicating that the method identifies the outlier as depicted in Fig.~\ref{fig: N-version programming method}.}
    \label{fig: Experimental results of the N-version programming method}
\end{figure}

\begin{table}[htb]
    \centering
    \begin{tabular}{|l||c|c|c|}
    \hline
       \multirow{2}{*}{Method} &  No.~of  & No.~of  & No. of   \\ 
        &  1st places  &  2nd places &  3rd places  \\ \hline \hline
       Linear       & 1& 4& 58   \\  \hline
       Richardson   & 0& 0& 12 \\ \hline 
       Exponential  & 22& 54& 2     \\ \hline
       Consistency-based    & \multirow{2}{*}{60} & \multirow{2}{*}{21} & \multirow{2}{*}{1}   \\ 
       method  &    &  &  \\ \hline 
    \end{tabular}
    \caption{Numerical experimental results of the consistency-based method in Sec.~\ref{sec: consistency-based method}. The columns ``No.~of 1st–4th places'' present, for each method, the frequency with which it obtained each respective rank based on TVD across 10 × 10 = 100 experimental runs.}
    \label{tab: Experimental results of the consistency-based method}
\end{table}

Here, we present the results of the experiments conducted for the proposed methods.
Figure~\ref {fig: Experimental results of the N-version programming method} summarizes the ranking of the QEM methods selected by the N-version programming approach in terms of TVD.
The Trotter number $M$ was varied from 5 to 10. For each fixed $M$, 100 trials are conducted, and the rank of the QEM method chosen by the proposed approach, based on TVDs among all candidate methods, is recorded. The vertical axis indicates the number of occurrences for each rank. In this experiment, the QEM method most frequently selected by the proposed approach is the second-ranked one, while the fourth-ranked (i.e., the worst-performing) method has never been chosen. This result indicates that, although the proposed method does not necessarily select the best-performing QEM, it consistently avoids poor (outlier) methods. This demonstrates that the anticipated behavior shown in Fig.~\ref{fig: N-version programming method} has been successfully observed. We remark that the method considered an outlier is always the polynomial-exponential extrapolation method.

Table~\ref{tab: Experimental results of the consistency-based method} presents the results of the experiment for the consistency-based method with $M=10$. In this numerical experiment, only linear, second-order Richardson, and exponential extrapolation are considered because the polynomial-exponential, which had been an outlier, can be excluded. The columns ``No.~of 1st–3rd places'' show how often each method attains each respective rank, based on TVDs computed over 
$10\times 10$ experimental runs. The consistency-based method achieves the first rank in 60 out of 100 trials, demonstrating that it is significantly more likely to achieve higher accuracy than the other QEM methods. Moreover, even when it does not rank first, our method ranks second in 20 trials and ranks last only once. In contrast, among the conventional QEM methods, specific approaches, such as the linear extrapolation method, perform considerably worse, ranking last in 58 trials.

\section{Conclusion and discussions}
In this work, we propose two methods to improve error-mitigation performance, inspired by software engineering for reliable classical computing. The first method is the N-version programming method, which compares the multiple error-mitigated distributions and excludes the atypical distribution. The second method is the consistency-based method. This method computes the variances of the error-mitigated results across multiple QEM methods and selects the result with the smallest variance as the most consistent; we also apply this method to each bin of the probability distribution. We confirm that the N-version programming method successfully excludes the most faulty error-mitigated distribution, and that the consistency-based method for probability distributions can achieve the most accurate error-mitigated distribution in the majority of cases.

While we focus on zero-noise extrapolation to demonstrate our protocol, our method can be naturally extended to other QEM methods, such as probabilistic error cancellation (PEC) \cite{temmeErrorMitigationShortDepth2017}. It may be interesting to include the PEC or the PEC-extrapolation combination for obtaining the error-mitigated distribution~\cite{mari2021extending,sun2021mitigating}, and to examine how the performance of our method changes with the accuracy of the noise characterization required for the PEC implementation.

In addition, the application of this method in the error-corrected regime is worth exploring. While our protocol has been demonstrated in the NISQ setup, i.e., any QEC functionality is not assumed, the interplay between the quantum error correction/detection and QEM has been recently studied for high-accuracy quantum computing in the early fault-tolerant quantum computing era~\cite{suzukiQuantumErrorMitigation2022,piveteau2021error,lostaglio2021error,tsubouchi2023virtual,xiong2020sampling}. Therefore, delving further into the optimization of our method and seeking suitable software engineering methods that consider the QEC functionality is of practical interest. 

\begin{acknowledgments}
The authors would like to thank Takayoshi Shiraki, Masayuki Oda, and Hidetoshi Sawada for valuable discussions and support with numerical simulations. 
\end{acknowledgments}

\bibliographystyle{apsrev4-2}
\bibliography{main}

@article{endoHybridQuantumClassicalAlgorithms2021,
author = {Endo ,Suguru and Cai ,Zhenyu and Benjamin ,Simon C. and Yuan ,Xiao},
title = {Hybrid Quantum-Classical Algorithms and Quantum Error Mitigation},
journal = {J. Phys. Soc. Jpn.},
volume = {90},
number = {3},
pages = {032001},
year = {2021},
doi = {10.7566/JPSJ.90.032001},
url = {https://doi.org/10.7566/JPSJ.90.032001}
}

@article{endoMitigatingAlgorithmicErrors2019,
  title = {Mitigating Algorithmic Errors in a {{Hamiltonian}} Simulation},
  author = {Endo, Suguru and Zhao, Qi and Li, Ying and Benjamin, Simon and Yuan, Xiao},
  year = {2019},
  month = jan,
  journal = {Phys. Rev. A},
  volume = {99},
  number = {1},
  pages = {012334},
  doi = {10.1103/PhysRevA.99.012334}
}

@article{endoPracticalQuantumError2018,
  title = {Practical Quantum Error Mitigation for Near-Future Applications},
  author = {Endo, Suguru and Benjamin, Simon C. and Li, Ying},
  journal = {Phys. Rev. X},
  volume = {8},
  issue = {3},
  pages = {031027},
  numpages = {21},
  year = {2018},
  month = {Jul},
  publisher = {American Physical Society},
  doi = {10.1103/PhysRevX.8.031027},
  url = {https://link.aps.org/doi/10.1103/PhysRevX.8.031027}
}

@article{kimEvidenceUtilityQuantum2023,
  title = {Evidence for the Utility of Quantum Computing before Fault Tolerance},
  author = {Kim, Youngseok and Eddins, Andrew and Anand, Sajant and Wei, Ken Xuan and {van den Berg}, Ewout and Rosenblatt, Sami and Nayfeh, Hasan and Wu, Yantao and Zaletel, Michael and Temme, Kristan and Kandala, Abhinav},
  year = {2023},
  month = jun,
  journal = {Nature},
  volume = {618},
  number = {7965},
  pages = {500--505},
  publisher = {Nature Publishing Group},
  issn = {1476-4687},
  doi = {10.1038/s41586-023-06096-3},
  copyright = {2023 The Author(s)},
  langid = {english}
}

@article{lloydUniversalQuantumSimulators1996,
  title = {Universal {{Quantum Simulators}}},
  author = {Lloyd, Seth},
  year = {1996},
  month = aug,
  journal = {Science},
  volume = {273},
  number = {5278},
  pages = {1073--1078},
  issn = {0036-8075, 1095-9203},
  doi = {10.1126/science.273.5278.1073},
  copyright = {{\copyright} 1996 American Association for the Advancement of Science},
  langid = {english}
}

@book{nielsenQuantumComputationQuantum2012,
    place={Cambridge},
    title={Quantum Computation and Quantum Information: 10th Anniversary Edition},
    publisher={Cambridge University Press},
    author={Nielsen, Michael A. and Chuang, Isaac L.},
    doi = {10.1017/CBO9780511976667},
    year={2010},
}

@article{suzukiGeneralizedTrotterFormula1976,
  title = {Generalized {{Trotter}}'s Formula and Systematic Approximants of Exponential Operators and Inner Derivations with Applications to Many-Body Problems},
  author = {Suzuki, Masuo},
  year = {1976},
  month = jun,
  journal = {Commun.Math. Phys.},
  volume = {51},
  number = {2},
  pages = {183--190},
  issn = {1432-0916},
  doi = {10.1007/BF01609348},
  langid = {english}
}

@article{suzukiQuantumErrorMitigation2022,
  title = {Quantum {{Error Mitigation}} as a {{Universal Error Reduction Technique}}: {{Applications}} from the {{NISQ}} to the {{Fault-Tolerant Quantum Computing Eras}}},
  shorttitle = {Quantum {{Error Mitigation}} as a {{Universal Error Reduction Technique}}},
  author = {Suzuki, Yasunari and Endo, Suguru and Fujii, Keisuke and Tokunaga, Yuuki},
  year = {2022},
  month = mar,
  journal = {PRX Quantum},
  volume = {3},
  number = {1},
  pages = {010345},
  issn = {2691-3399},
  doi = {10.1103/PRXQuantum.3.010345},
  langid = {english}
}

@article{temmeErrorMitigationShortDepth2017,
  title = {Error {{Mitigation}} for {{Short-Depth Quantum Circuits}}},
  author = {Temme, Kristan and Bravyi, Sergey and Gambetta, Jay M.},
  year = {2017},
  month = nov,
  journal = {Phys. Rev. Lett.},
  volume = {119},
  number = {18},
  pages = {180509},
  doi = {10.1103/PhysRevLett.119.180509}
}

@article{muqeetQUIETToolforSampling-BasedQuantumNoiseErrorMitigation,
  author={Muqeet, Asmar and Ali, Shaukat and Arcaini, Paolo},
  journal={IEEE Software}, 
  title={Tool: QUIET: A Tool for Sampling-Based Quantum Noise Error Mitigation}, 
  year={2025},
  volume={42},
  number={6},
  pages={28-34},
  doi={10.1109/MS.2025.3532106}
}

@misc{liu2025quantum,
      title={Quantum Error Mitigation for Sampling Algorithms}, 
      author={Kecheng Liu and Zhenyu Cai},
      year={2025},
      eprint={2502.11285},
      archivePrefix={arXiv},
      primaryClass={quant-ph},
      url={https://arxiv.org/abs/2502.11285}, 
}

@misc{kanno2023quantumselectedconfigurationinteractionclassical,
      title={Quantum-Selected Configuration Interaction: classical diagonalization of Hamiltonians in subspaces selected by quantum computers}, 
      author={Keita Kanno and Masaya Kohda and Ryosuke Imai and Sho Koh and Kosuke Mitarai and Wataru Mizukami and Yuya O. Nakagawa},
      year={2023},
      eprint={2302.11320},
      archivePrefix={arXiv},
      primaryClass={quant-ph},
      url={https://arxiv.org/abs/2302.11320}, 
}

@misc{qiskit2024,
      title={Quantum computing with {Q}iskit},
      author={Javadi-Abhari, Ali and Treinish, Matthew and Krsulich, Kevin and Wood, Christopher J. and Lishman, Jake and Gacon, Julien and Martiel, Simon and Nation, Paul D. and Bishop, Lev S. and Cross, Andrew W. and Johnson, Blake R. and Gambetta, Jay M.},
      year={2024},
      doi={10.48550/arXiv.2405.08810},
      eprint={2405.08810},
      archivePrefix={arXiv},
      primaryClass={quant-ph}
}

@article{childsTheoryTrotterError2021,
  title = {Theory of {{Trotter Error}} with {{Commutator Scaling}}},
  author = {Childs, Andrew M. and Su, Yuan and Tran, Minh C. and Wiebe, Nathan and Zhu, Shuchen},
  year = {2021},
  month = feb,
  journal = {Phys. Rev. X},
  volume = {11},
  number = {1},
  pages = {011020},
  publisher = {American Physical Society},
  doi = {10.1103/PhysRevX.11.011020}
}

@INPROCEEDINGS{NVP1995,
  author={Liming Chen and Avizienis, A.},
  booktitle={Twenty-Fifth International Symposium on Fault-Tolerant Computing, 1995, ' Highlights from Twenty-Five Years'.}, 
  title={N-VERSION PROGRAMMINC: A FAULT-TOLERANCE APPROACH TO RELlABlLlTY OF SOFTWARE OPERATlON}, 
  year={1995},
  volume={},
  number={},
  pages={113--},
  doi={10.1109/FTCSH.1995.532621}
}

@INPROCEEDINGS{SaitoRAQuS2024,
  author={Saito, Shinobu and Endo, Suguru and Suzuki, Yasunari},
  booktitle={2024 IEEE 35th International Symposium on Software Reliability Engineering Workshops (ISSREW)}, 
  title={Towards N-version Quantum Software Systems for Reliable Classical-Quantum Computing}, 
  year={2024},
  volume={},
  number={},
  pages={119--120},
  doi={10.1109/ISSREW63542.2024.00064}
}

@article{computing2006architectural,
    author = {Steve R. White and James E. Hanson and Ian Whalley and David M. Chess and Alla Segal and Jeffrey O. Kephart},
    title ={Autonomic computing: Architectural approach and prototype},
    journal = {Integrated Computer-Aided Engineering},
    volume = {13},
    number = {2},
    pages = {173-188},
    year = {2006},
    doi = {10.3233/ICA-2006-13206},
    URL = {https://journals.sagepub.com/doi/abs/10.3233/ICA-2006-13206}
}

@article{Nakagawa2024adaptQSCI,
author = {Nakagawa, Yuya O. and Kamoshita, Masahiko and Mizukami, Wataru and Sudo, Shotaro and Ohnishi, Yu-ya},
title = {ADAPT-QSCI: Adaptive Construction of an Input State for Quantum-Selected Configuration Interaction},
journal = {J. Chem. Theory Comput.},
volume = {20},
number = {24},
pages = {10817-10825},
year = {2024},
doi = {10.1021/acs.jctc.4c00846},
url = {https://doi.org/10.1021/acs.jctc.4c00846},
}

@INPROCEEDINGS{Tiron2020DigitalZeroNoiseExtrapolationGateFoldingPaper,
  author={Giurgica-Tiron, Tudor and Hindy, Yousef and LaRose, Ryan and Mari, Andrea and Zeng, William J.},
  booktitle={2020 IEEE International Conference on Quantum Computing and Engineering (QCE)}, 
  title={Digital zero noise extrapolation for quantum error mitigation}, 
  year={2020},
  volume={},
  number={},
  pages={306-316},
  doi={10.1109/QCE49297.2020.00045}
}

@article{mitiq2022Quantum,
  title     = {Mitiq: A software package for error mitigation on noisy quantum computers},
  author    = {Ryan LaRose and Andrea Mari and Sarah Kaiser and Peter J. Karalekas and Andre A. Alves and Piotr Czarnik and Mohamed El Mandouh and Max H. Gordon and Yousef Hindy and Aaron Robertson and Purva Thakre and Misty Wahl and Danny Samuel and Rahul Mistri and Maxime Tremblay and Nick Gardner and Nathaniel T. Stemen and Nathan Shammah and William J. Zeng},
  journal   = {Quantum},
  year      = {2022},
  month     = {Aug},
  doi       = {10.22331/q-2022-08-11-774},
  url       = {https://doi.org/10.22331/q-2022-08-11-774},
  publisher = {Verein zur Forderung des Open Access Publizierens in den Quantenwissenschaften},
  volume    = {6},
  pages     = {774},
}

@article{preskill2018quantum,
  title={Quantum computing in the NISQ era and beyond},
  author={Preskill, John},
  journal={Quantum},
  volume={2},
  pages={79},
  year={2018},
  publisher={Verein zur F{\"o}rderung des Open Access Publizierens in den Quantenwissenschaften},
  url={https://quantum-journal.org/papers/q-2018-08-06-79/}
}

@article{preskill2025beyond,
    author = {Preskill, John},
    title = {Beyond NISQ: The Megaquop Machine},
    year = {2025},
    issue_date = {September 2025},
    publisher = {Association for Computing Machinery},
    address = {New York, NY, USA},
    volume = {6},
    number = {3},
    journal = {ACM Transactions on Quantum Computing},
    month = apr,
    pages = {18},
    numpages = {7},
    url = {https://doi.org/10.1145/3723153},
    doi = {10.1145/3723153},
}

@misc{eisert2025mind,
    title={Mind the gaps: The fraught road to quantum advantage}, 
    author={Jens Eisert and John Preskill},
    year={2025},
    eprint={2510.19928},
    archivePrefix={arXiv},
    primaryClass={quant-ph},
    url={https://arxiv.org/abs/2510.19928},
}

@article{devitt2013quantum,
  title={Quantum error correction for beginners},
  author={Devitt, Simon J and Munro, William J and Nemoto, Kae},
  journal={Rep. Prog. Phys.},
  volume={76},
  number={7},
  pages={076001},
  year={2013},
  publisher={IOP Publishing},
  url={https://iopscience.iop.org/article/10.1088/0034-4885/76/7/076001}
}

@book{lidar2013quantum,
  title={Quantum error correction},
  author={Lidar, Daniel A and Brun, Todd A},
  year={2013},
  publisher={Cambridge university press},
  url={https://doi.org/10.1017/CBO9781139034807}
}

@article{cai2023quantum,
  title = {Quantum error mitigation},
  author = {Cai, Zhenyu and Babbush, Ryan and Benjamin, Simon C. and Endo, Suguru and Huggins, William J. and Li, Ying and McClean, Jarrod R. and O'Brien, Thomas E.},
  journal = {Rev. Mod. Phys.},
  volume = {95},
  issue = {4},
  pages = {045005},
  numpages = {37},
  year = {2023},
  month = {Dec},
  publisher = {American Physical Society},
  doi = {10.1103/RevModPhys.95.045005},
  url = {https://link.aps.org/doi/10.1103/RevModPhys.95.045005}
}

@article{li2017efficient,
  title = {Efficient Variational Quantum Simulator Incorporating Active Error Minimization},
  author = {Li, Ying and Benjamin, Simon C.},
  journal = {Phys. Rev. X},
  volume = {7},
  issue = {2},
  pages = {021050},
  numpages = {14},
  year = {2017},
  month = {Jun},
  publisher = {American Physical Society},
  doi = {10.1103/PhysRevX.7.021050},
  url = {https://link.aps.org/doi/10.1103/PhysRevX.7.021050}
}

@misc{cai2021practical,
  title={A Practical Framework for Quantum Error Mitigation}, 
  author={Zhenyu Cai},
  year={2023},
  eprint={2110.05389},
  archivePrefix={arXiv},
  primaryClass={quant-ph},
  url={https://arxiv.org/abs/2110.05389}, 
}

@article{mari2021extending,
  title = {Extending quantum probabilistic error cancellation by noise scaling},
  author = {Mari, Andrea and Shammah, Nathan and Zeng, William J.},
  journal = {Phys. Rev. A},
  volume = {104},
  issue = {5},
  pages = {052607},
  numpages = {12},
  year = {2021},
  month = {Nov},
  publisher = {American Physical Society},
  doi = {10.1103/PhysRevA.104.052607},
  url = {https://link.aps.org/doi/10.1103/PhysRevA.104.052607}
}

@article{sun2021mitigating,
  title = {Mitigating Realistic Noise in Practical Noisy Intermediate-Scale Quantum Devices},
  author = {Sun, Jinzhao and Yuan, Xiao and Tsunoda, Takahiro and Vedral, Vlatko and Benjamin, Simon C. and Endo, Suguru},
  journal = {Phys. Rev. Appl.},
  volume = {15},
  issue = {3},
  pages = {034026},
  numpages = {23},
  year = {2021},
  month = {Mar},
  publisher = {American Physical Society},
  doi = {10.1103/PhysRevApplied.15.034026},
  url = {https://link.aps.org/doi/10.1103/PhysRevApplied.15.034026}
}

@article{piveteau2021error,
  title = {Error Mitigation for Universal Gates on Encoded Qubits},
  author = {Piveteau, Christophe and Sutter, David and Bravyi, Sergey and Gambetta, Jay M. and Temme, Kristan},
  journal = {Phys. Rev. Lett.},
  volume = {127},
  issue = {20},
  pages = {200505},
  numpages = {6},
  year = {2021},
  month = {Nov},
  publisher = {American Physical Society},
  doi = {10.1103/PhysRevLett.127.200505},
  url = {https://link.aps.org/doi/10.1103/PhysRevLett.127.200505}
}

@article{lostaglio2021error,
  title = {Error Mitigation and Quantum-Assisted Simulation in the Error Corrected Regime},
  author = {Lostaglio, M. and Ciani, A.},
  journal = {Phys. Rev. Lett.},
  volume = {127},
  issue = {20},
  pages = {200506},
  numpages = {6},
  year = {2021},
  month = {Nov},
  publisher = {American Physical Society},
  doi = {10.1103/PhysRevLett.127.200506},
  url = {https://link.aps.org/doi/10.1103/PhysRevLett.127.200506}
}

@article{tsubouchi2023virtual,
  title = {Virtual quantum error detection},
  author = {Tsubouchi, Kento and Suzuki, Yasunari and Tokunaga, Yuuki and Yoshioka, Nobuyuki and Endo, Suguru},
  journal = {Phys. Rev. A},
  volume = {108},
  issue = {4},
  pages = {042426},
  numpages = {10},
  year = {2023},
  month = {Oct},
  publisher = {American Physical Society},
  doi = {10.1103/PhysRevA.108.042426},
  url = {https://link.aps.org/doi/10.1103/PhysRevA.108.042426}
}

@article{xiong2020sampling,
  author={Xiong, Yifeng and Chandra, Daryus and Ng, Soon Xin and Hanzo, Lajos},
  journal={IEEE Access}, 
  title={Sampling Overhead Analysis of Quantum Error Mitigation: Uncoded vs. Coded Systems}, 
  year={2020},
  volume={8},
  number={},
  pages={228967-228991},
  doi={10.1109/ACCESS.2020.3045016}
}

\appendix
\section{Trotterization}\label{sec: Trotterization}
Here, we briefly explain the Trotterization algorithm \cite{suzukiGeneralizedTrotterFormula1976,lloydUniversalQuantumSimulators1996,childsTheoryTrotterError2021}, which approximates an exponential of a sum of operators, such as a time evolution operator, by a product of elementary exponentials.
It is used as a subroutine of many quantum algorithms, including Hamiltonian simulation \cite{lloydUniversalQuantumSimulators1996} and phase estimation \cite{nielsenQuantumComputationQuantum2012}.
In the following, we explain the first-order Trotterization algorithm using a time evolution operator as an example. Let $H=\sum_i^LH_i$, where each $H_i$ acts on a constant number of qubits. The first-order Trotterization of the time evolution operator $e^{-iHt}$ is given by
\begin{equation}
    S_1(t) = \prod_j e^{-iH_jt}.
\end{equation}
For the Hamiltonian simulation using a quantum circuit, we implement the first-order Trotterization $S_1\pqty{t/M}$ for $M$ times.
$M$ is called the Trotter number, and it controls the algorithmic error (Trotter error). Ref.~\cite{lloydUniversalQuantumSimulators1996} has shown that
\begin{equation}
\begin{aligned}
e^{-iHt} - \{S_1(t/M)\}^M= \frac{t^2}{2M}\sum_{i<j} [H_i, H_j] + \sum_{k}^{\infty} E_k
\end{aligned},
\end{equation}
where $\| E_k \| \leq M \|H t/M \|^k/k!$ for the operator norm $\| \cdot\|$.  If the quantum circuits are error-free, the algorithmic error can be arbitrarily suppressed as the Trotter number $M$ increases.

\section{Software used in numerical experiments}\label{sec: Software table}
\begin{table}[tb]
    \centering
    \begin{tabular}{|l|l|}
    \hline
       OS & Ubuntu 24.04 LTS \\ \hline 
       Python  & 3.12.9\\ \hline 
       pip  & 25.2 \\ \hline 
        qiskit  &   1.4.0\\ \hline 
        qiskit-aer-gpu &  0.15.1\\ \hline 
        qiskit-ibm-runtime &  0.36.0\\ \hline 
        simulator backend & \texttt{FakeMarrakesh}\\ \hline
        mitiq & 0.47.0\\ \hline 
    \end{tabular}
    \caption{Software used for our numerical experiments.}
    \label{tab: software used for the experiments.}
\end{table}
The software used in our numerical experiments is summarized in Tab.~\ref{tab: software used for the experiments.}.

\section{Error amplification}\label{sec: Error amplification}

In this section, we present the error-amplification procedure used in Sec.~\ref{sec: Experimental setup and procedure}. 
The procedure is described in Procedure~\ref{procedure: Error amplification via gate folding}. For a given unitary circuit $U$, "AmplifyError" effectively amplifies the error in the circuit by a scale factor $\lambda$. This is achieved by employing the gate folding technique proposed in Ref.~\cite{Tiron2020DigitalZeroNoiseExtrapolationGateFoldingPaper}, in which redundant gates are inserted before and after each original gate to effectively increase the circuit’s error rate by the factor $\lambda$. However, in our implementation, we perform an additional post-processing step to remove all $\text{(SX)}^\dagger$ gates after applying gate folding. This modification is necessary because, when the circuit shown in Fig.~\ref{fig: example_of_Trotterized_circuit} is transpiled for the \texttt{FakeMarrakesh} backend, SX gates appear as part of the compiled circuit. If gate folding is applied to such a circuit, the resulting folded circuit inevitably contains $\text{(SX)}^\dagger$ gates. Since the \texttt{FakeMarrakesh} backend does not support the $\text{(SX)}^\dagger$ gate as a native operation, circuits containing this gate cannot be executed directly. Therefore, we removed all $\text{(SX)}^\dagger$ gates from the folded circuits before execution. As a consequence of this removal, the "AmplifyError" procedure sacrifices accuracy in the effective error-rate scaling. 

\floatname{algorithm}{Procedure}
\begin{algorithm}[H]
\begin{algorithmic}[1]
\Procedure{AmplifyError}{$\lambda$, $U$}
    \State $\circ$ Assume $\lambda$ is an odd integer.
    \State $\circ\; U_{\lambda}:$ Empty circuit.
    \State $\circ\; g^\text{list}:$ List of the gates in $U$ (excluding the measurement gates).
    \For{$g$ in $g^\text{list}$}
        \State $\circ\; U_{\lambda}\leftarrow(gg^\dagger)^{\lambda-1}g U_{\lambda}$.
    \EndFor
    \State $\circ$ Remove the $\text{(SX)}^\dagger$ gates in $U_{\lambda}$. This is due to the constraints in the \texttt{FakeMarrakesh} backend, which does not support the $\text{(SX)}^\dagger$ gates.
    \\
    \Return $U_{\lambda}$.
\EndProcedure
\caption{Error amplification via gate folding.}
\label{procedure: Error amplification via gate folding}
\end{algorithmic}
\end{algorithm}
\floatname{algorithm}{Algorithm}

\end{document}